\documentstyle[twocolumn, aps, epsf, overcite]{revtex}
\def\bea{\begin{eqnarray}}
\def\eea{\end{eqnarray}}
\begin{document}

\draft
\title{\bf  Landau Level Mixing and Skyrmion Stability in Quantum Hall Ferromagnets }

\author{I. Mihalek and H.A. Fertig}
\address{Department of Physics and Astronomy,
University of Kentucky,  Lexington, KY 40506-0055}
\date{\today}
\maketitle

\begin{abstract}
  We present a Hartree-Fock study  that incorporates the effects of 
Landau level mixing and screening due to filled levels into 
the computation of energies and states of quasiparticles in  quantum
Hall ferromagnets.
We use it  to construct a phase diagram for skyrmion
stability as a function of magnetic field and Zeeman coupling strengths.
We find that Landau level mixing tends to favor spin-polarized
quasiparticles, while finite thickness corrections favor skyrmions.
Our studies show that skyrmion stability in high Landau levels
is very sensitive to the  way in which electron-electron
interactions are modified by finite thickness, and indicate
that it is crucial to use models with realistic short distance
behavior to get qualitatively correct results.
We find that recent experimental evidence for skyrmions
in higher Landau levels cannot be explained within our model.
\end{abstract}

\section{Introduction}
It has been recognized
 that exotic magnetic excitations
known as skyrmions may exist \cite{lee,sondhi,herb}
in a two-dimensional electron gas in a strong homogeneous
magnetic field (quantum Hall system) 
 near spin
polarized groundstates. 
 These are excitations  of a two-dimensional spontaneous
ferromagnet, the physics of which is relevant to  this 
 system (despite the
presence of a strong magnetic field),  because of 
the small Land\'e $g$-factor in GaAs systems (where
most experiments take place), which makes the Zeeman coupling
very  small compared to other energy scales
(Coulomb interaction, cyclotron energy) in the problem.
Skyrmions  are  spin configurations
with a non-trivial winding number (Pontryagin index).  They were first
discussed in the context of  four-dimensional field theories 
\cite{skyrme,rajaraman}, and
were later recognized as states occurring in the non-linear sigma model
description of two-dimensional ferromagnets \cite{polyakov}.
 For filling factors $\nu \equiv N/N_{\phi}$ close
to one ($N$ is the number of electrons and $N_{\phi}$
the number of magnetic flux quanta penetrating the system),
these turn out to be the lowest energy quasiparticles 
under typical experimental circumstances.
Skyrmions can thus be introduced  into the
groundstate by adding or removing
charge from the system.
\cite{herb}. 

Experimentally, the case for the existence of skyrmions
in a system close to  $\nu=1$ is  quite strong.  NMR experiments
show a degrading of the spin polarization
with deviation of filling factor from one \cite{barret} that is 
in remarkably good agreement with Hartree-Fock
theory \cite{herb,brey}. The quasiparticle
spin
 measured in transport experiments \cite{eis}
are also reasonably well accounted for by Hartree-Fock
calculations \cite{herb}.  Electromagnetic absorption
experiments \cite{aifer} further support that doping
away from $\nu=1$ injects skyrmions into the system.

In weaker magnetic fields, near filling factor $\nu=3$,
early experiments \cite{eis} suggested that spin-polarized 
quasiparticles are lower in energy than skyrmions,
so  the effects seen near $\nu=1$ would
not be present at higher filling.  This is
consistent with calculations of skyrmion energies
near $\nu=3$ that include finite thickness
corrections \cite{cooper,herb2}, which indicate
that skyrmions will be present only at much
smaller Zeeman couplings than realized in typical
experiments. 
 The size of the skyrmion
may be quantified by a number  $K$,  the difference in the
spin component $S_z$ between the skyrmion and the
spin-polarized quasiparticle. Because of the necessarily  small
Zeeman coupling, stable skyrmions close to  $\nu=3$
have large values of  $K$.  They also become   unstable with
respect to spin-polarized quasiparticles at a 
finite  value of $K$.
 (For
$\nu=1$,  $K \rightarrow 0$
as the Zeeman coupling reaches the maximum value for 
which the system supports skyrmions; i.e., the skyrmion
state smoothly goes into the spin-polarized quasiparticle
state.)  For a two-dimensional electron gas (2DEG) with
width of about 2$\ell$, where $\ell=\sqrt{\hbar c/ eB}$
is the magnetic length, the minimum $K$ expected  \cite{cooper} 
is approximately
4.  

Recently however, NMR experiments \cite{song} have uncovered evidence that some
anomalous degrading of spin polarization $does$ occur
as one dopes away from $\nu=3$ at relatively
high Zeeman couplings.  These experiments
further indicate that the number of overturned spins
per quasiparticle is quite small, $K \sim 1$.
The simple models usually considered \cite{cooper,herb2}
are inconsistent with this, and one is naturally
led to inquire as to what other ingredients
might change the critical Zeeman coupling and
smallest $K$ observable near $\nu=3$.  Two possible
answers are Landau level mixing and screening by
filled Landau levels.  It should be noted that
these are not distinct effects: screening
by filled Landau levels occurs because they
may admix high (unoccupied) Landau levels to smooth
fluctuations due to external potentials and/or inhomogeneous
electron densities in partially filled levels.  
Conversely, the states which may be used 
for Landau level mixing in a partially filled
level are limited to those that are not occupied
by electrons in other levels,  due to Pauli exclusion.
Thus, a correct treatment of either screening
or Landau level mixing near $\nu=3$
must include  both  these effects.

In this work, we present a method by which
these may be incorporated into the Hartree-Fock
description of skyrmion states. 

Our principal conclusions may be summarized as
follows:
 (i) For skyrmions near $\nu=1$,
Landau level mixing tends to lower the quasiparticle
energy, although not enough to quantitatively
explain the activation energies seen in
experiment \cite{eis}.  Introduction of a 
finite width of one magnetic length lowers
the energy of the skyrmion by approximately
40\%, and inclusion of Landau
level mixing lowers the energy by approximately
another 20\% for $\hbar \omega_c/(e^2 /\kappa \ell)
\approx 0.6$, where $\omega_c=eB/m^*c$ is the
cyclotron frequency of the electrons, and 
$\kappa$ is the dielectric constant of the
host crystal.  The resulting quasiparticle
gap is approximately a factor of 2 larger
than what is found in experiment \cite{eis}.
This result agrees qualitatively with that
of another study of Landau level mixing effects
on $\nu=1$ skyrmions \cite{melik}.  
(ii) For $\nu=1$, Landau level mixing tends to 
suppress quasihole-like skyrmions (i.e., lowering
the maximum Zeeman coupling for which they are
stable), while enhancing the stability of
quasielectron (anti-)skyrmions.  
(iii) For $\nu=3$ and  higher,
 we find that a sufficiently realistic
model  of the effective electron-electron
interaction, as modified by the finite thickness
of the electronic wavefunctions, is necessary
to obtain reliable results.  Use of a simple
potential due to Zhang and Das Sarma \cite{zhang}
grossly overestimates the stability of $\nu=3$
skyrmions; more realistic potentials \cite{cooper,herb2}
allow skyrmions only for very small Zeeman
couplings.  
(iv) Screening and Landau level mixing for $\nu=3$ and $\nu=5$
tend to lower the energy of spin-polarized quasiparticles
more than that of skyrmions, making the latter even
less stable.
(v) The results of Ref. \citen{song} cannot be
understood solely on the basis of Hartree-Fock 
states for skyrmions.  

The remainder of this article is organized as
follows.  In Section II below, we discuss the
method used to allow screening and Landau level
mixing to be included in our calculations.
Section III gives details of our results,
and we conclude with a summary of our findings
in Section IV.

\section{Hartree-Fock Method with Landau Level Mixing}

 Most previous Hartree-Fock studies  of skyrmions have relied
on Landau level representation of the single particle 
states \cite{herb}. 
We choose instead to construct the wavefunctions in real space.
This enables us to include in the model  Landau level mixing occurring
in  weak magnetic fields, without having to expand over the large
number of Landau levels necessary in the former approach.
We thus trade the calculational
convenience of working with the functions given in  closed analytic
form (Landau levels) for a  closer description of the single-particle
 states by representing them on a real-space grid.

In this calculation we aim to model
all   the participating particles, including the ones in the filled levels. 
Our Hartree-Fock wavefunction  is a  Slater determinant composed of
single-particle  states which have
$L_z\pm S_z$ as a good quantum number \cite{herb} but whose radial form is to be determined
self-consistently:
\begin{eqnarray}\label{trial}
| \Psi_{skyrmion} \rangle = \prod_{i,m} \gamma_{im}^\dagger |0\rangle 
  \nonumber \\
\langle \vec r, \sigma| \gamma_{im}^\dagger |0\rangle =  
\left[
\begin{array}{c}
 f_{im}(r) \\
 g_{im}(r)e^{\pm i\theta }
\end{array}
\right] e^{im\theta }.
\end{eqnarray} 
Here $r$ and $\theta$  are polar coordinates,  $m$ is the angular  momentum
quantum number, and $i$ labels different states of the same $m$. The sign
 $\pm$ corresponds to two  
families of solutions, $+$ for  antiskyrmion (or quasielectron spin structured
solution) and $-$ for skyrmion (quasihole).

 In very strong magnetic fields, the functions $f(r)$ and $g(r)$ 
take the  form expected for  Landau level states. 
 When the strength of the magnetic field  is lowered to bring the
ratio of cyclotron  and  Coulomb energy scales close to 1
, the form of the radial part relaxes toward some modified
form, as dictated by the interactions in the system.

Using the trial form of the wavefunction (\ref{trial}), 
 the many-body  Schr\"odinger equation
with the Hamiltonian
\begin{eqnarray} 
&&H= \frac{1}{2m} \int d^2r \sum_\sigma \Psi_\sigma^\dagger (\vec r)
|\frac {\hbar}{i} \vec \nabla-\frac {e}{c}\vec A|^2 \Psi _\sigma(\vec r) \nonumber\\
&+&\frac {1}{2}g\mu B \int  d^2r 
[ \Psi_\downarrow^\dagger (\vec r)\Psi_\downarrow (\vec r) -
\Psi_\uparrow^\dagger (\vec r)\Psi_\uparrow (\vec r)] \nonumber\\
&+&\frac {1}{2} \sum_{\sigma \sigma'} \int  d^2r  d^2r '
\Psi_\sigma^\dagger (\vec r)\Psi_{\sigma'} (\vec r') v(\vec r - \vec r')
\Psi_{\sigma'}^\dagger (\vec r')\Psi_\sigma (\vec r)
\end{eqnarray}
(where $\sigma$ denotes spin and $v(\vec r - \vec r')$ the 
Coulomb interaction),
upon variation with respect to the functions
$ f$ and $ g$, gives  a system of mean-field single-particle equations:

\begin{eqnarray} \label{master}
&&D_m(r) f_{im}(r)-\frac{1}{2}g\mu _BB f_{im}(r) \nonumber \\
&+& \int _{0}^{\infty }r'\,dr'V^H(r,r')[\rho (r')-\rho _0] f_{im}(r)\nonumber \\
&-&\int _{0}^{\infty }r'\,dr'\sum _{m'}V^{ex}_{m-m'}(r,r')
\rho _{m'}^{\uparrow \uparrow }(r',r) f_{im}(r') \nonumber \\
&-&\int _{0}^{\infty }r'\,dr'\sum _{m'}V^{ex}_{m-m'}(r,r')
\rho _{m'}^{\downarrow \uparrow }(r',r) g_{im\pm 1}(r') \nonumber \\
&=& \epsilon_i  f_{im}(r)
\end{eqnarray}
together with the analogous equation for the function $g(r)$. 
Here is a dictionary of the notation accompanying  Eq.
\ref{master}:
 the operator $D_m$ is given by
\begin{eqnarray}
D_m &=& -\frac {\hbar ^2}{2m^*}[\frac {1}{r}\frac {d}{dr}r\frac {d}{dr}
-\frac {m^2}{r^2}] \nonumber \\
&-& m\frac {\hbar \omega _c}{2}  
+\frac {(m^*)^2}{4}\frac {\omega _c^2r^2}{2m^*}
\end{eqnarray}
with $\omega _c = \frac {eB}{m^*c}$, $B$   the magnitude of the
external magnetic field, and  $ m^*$ the effective mass of the
electron. $g$ is the Land\'e g-factor, and $\mu_B$ is the Bohr magneton.
$\rho$'s denote  generalized densities:
\begin{eqnarray}
\rho _{m'}^{\uparrow \uparrow }(r',r) &=& f^*_{m'}(r')f_{m'}(r) \nonumber\\
\rho _{m'}^{\downarrow \downarrow }(r',r)&=&g^*_{m'\pm 1}(r')g_{m'\pm 1}(r) \nonumber\\
\rho _{m'}^{\downarrow \uparrow }(r',r)  &=&g^*_{m'\pm 1}(r')f_{m'}(r) \nonumber\\
\rho (r)  &=& \sum _{m'}
[ \rho _{m'}^{\uparrow \uparrow }(r,r) + \rho _{m'}^{\downarrow \downarrow }(r,r)],
\end{eqnarray}
and $\rho_0$ is the uniform background density.
$V^H$ and $V^{ex}$  are  the following integrals of the  Coulomb 
potential over the azimuthal variable: 
\begin{eqnarray} \label {potentials}
V^H(r,r')  &=&\int _0^{2\pi }d\theta \int _0^{2\pi } d\theta '
v(\vec r-\vec {r'}) \nonumber\\
V^{ex}_{m-m'}(r,r') &=& 
\int _0^{2\pi }d\theta \int _0^{2\pi } d\theta '
e^{i(m-m')(\theta -\theta ')}  v(\vec r-\vec {r'}),
\end{eqnarray}
 and, finally, $  \epsilon_i$ stands for the single-particle
Hartree-Fock energy.

 The finite width of the sample is modeled using the form of
 the in-plane potential due to  Cooper ~\cite{cooper}:
 we replace the Coulomb interaction 
$ v(\vec r-\vec {r'})$ in the Eq. ~(\ref{potentials}) by 

\begin{eqnarray}
v_C(r) = \int \int_ {-\infty}^{\infty}dzdz' 
  \frac{ e^{-(z'^2+ z^2)/2w^2}}{ 2\pi w^2}
  \frac {1} { \sqrt {r^2 + (z-z')^2 }     } .
\end{eqnarray}
The symbol $w$ denotes the width of system in the direction 
 perpendicular do the plane of the system,  and $z$, $z'$ are 
coordinates in that direction.

To handle the boundaries of the system, we assume the electron states with  
angular momentum  $ m>m_{max}$  have the ferromagnetic groundstate 
form (i.e. Landau levels with well defined 
spin).  The states with  $ m\leq m_{max}$ are explicitly included
in the calculation.  For $K$ not too large we find it is sufficient
to allow variations of  the states with $m$ of up to 30 for $\nu = 3$
and up to 50 for $\nu = 5$. 
In practice, including
boundary electrons from the 
states with $m$ between  31 and 100 ($\nu = 3$) and between    51 and 120 
($\nu = 5$) describes the effect of the system edge
 with precision matching 
the rest of the calculation.

Understanding Eq. (\ref{master}) as a system  
of coupled eigenproblems, we look
for  the self-consistent single particle solutions. (Discretization 
 will turn
each eigenequation into a matrix diagonalization problem which
can be handled using  standard methods.) 
The results we thus obtain will be largely
presented as comparisons  between energies  of the  
spin-polarized quasiparticle and 
energies of the corresponding skyrmion.

  To assess the energy of the skyrmion in the region of parameters where
it is not stable, we add to the Hamiltonian a term of the form 
 $\lambda(\hat S_z -S_0)^2$, $\hat S_z  $ being the spin operator,
and $\lambda$ a  tunable  parameter. This term favors  a state
with  total  spin  $S_0$, but  is insensitive  to the detailed form of the
wavefunctions. This allows the variational scheme to pick out the
lowest energy Slater determinant of the form given in Eq.(\ref{trial})
within the space of states with the same fixed value of $K$.

\section{Results}

 Based on the calculation  described in the previous section, we present
some of the results the method allows us to  obtain; we focus mainly
on the singly charged excitations  in the first three Landau levels.
Consideration of higher Landau levels is also possible, but computation
of the potential lookup tables becomes prohibitively expensive, and, as
the results so far indicate, leads to no new insight. In the following we
shall take the  unit of energy to be 
 $e^2/\kappa \ell$,  and the unitless   Zeeman splitting  to be 
$\tilde g = \frac{g\mu _BB}{e^2/\kappa \ell} $, where $e$ is the electron charge, 
$\kappa$ is the dielectric constant of  the host material, 
$\ell$ is the  magnetic length in the field $B$ and $\mu _B$ 
is the Bohr magneton.

\subsection {Skyrmion vs. polarized quasiparticle}

\begin{figure}[h]
\centerline{
  \epsfxsize=2.5in
  \epsffile{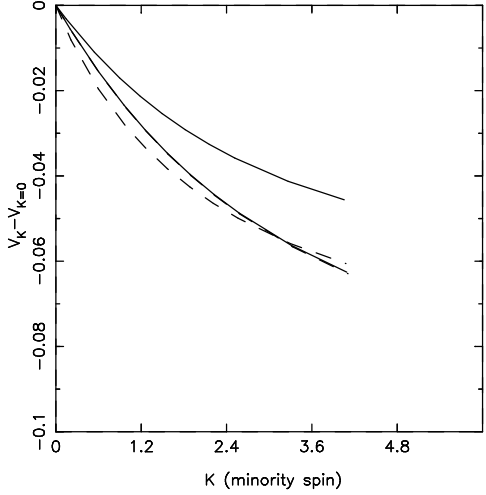}
}   
\vspace{0.2in}
\caption{ \label{lowest} 
Energy difference between skyrmion and polarized quasihole at $\nu$ =  1,
for  well width of $1.73 \ell$ and with $\tilde g \rightarrow 0 $.
The full lines correspond to the quasihole
and the dashed ones to the quasielectron case. For high cyclotron energies
($100.0 e^2/\kappa \ell$ ; the lower full curve) 
the results for the two types  of excitations overlap.
For low cyclotron energies ($ 0.8 e^2/\kappa \ell$;
 upper full line and non-overlapping dashed line) 
there is significant difference in the
behavior of the two, which reflects itself in the phase diagram, Fig. \ref{phased}. }
\end{figure}

  In Fig. \ref{lowest} we show  the energy difference between
the spin-polarized quasiparticle and the skyrmion of size $K$, $V_K-V_{K=0}$. 
This quantity is a pure interaction energy (i.e. Zeeman energy is not included),
and represents the energy gained or lost in deforming a spin-polarized
 quasiparticle into a skyrmion when Zeeman coupling is absent. Of particular importance
is the  slope (negative slopes indicate that  skyrmion is stable for
some value of $g$)  and the curvature (concave curves will support small-sized
skyrmions). For concave curves the largest Zeeman splitting that
supports skyrmions is
the negative of the initial slope of the curve \cite{herb2}.

For large cyclotron energies our results are essentially identical to those obtained
using the single Landau level method \cite{herb}. Note the quasielectron and
quasihole excitations are precisely degenerate in this case, due to
particle-hole symmetry. For smaller values of $\hbar\omega_c$, the two curves split;
surprisingly, the quasihole skyrmion is suppressed by Landau level mixing, whereas
the quasielectron skyrmion is enhanced. (The former result is in agreement
 with Ref. \citen {melik}).

\begin{figure}[h]
\centerline{
  \epsfxsize=2.5in
  \epsffile{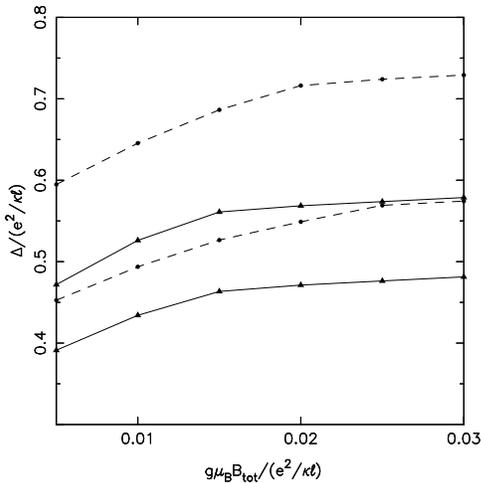}
}   
\vspace{0.2in}
\caption{ \label{gap}Skyrmion-antiskyrmion pair  
excitation gap for two different
well widths : $1.73 \ell$ (full line) and $1.0 \ell$ (dashed line). Upper lines
correspond to the cyclotron energy of $100.0e^2/\kappa \ell$ and lower
ones to $0.8 e^2/\kappa \ell$.}
\end{figure}

The energy gaps (Fig. \ref{gap}) 
which result from creation of skyrmion-antiskyrmion pairs when Landau level
 mixing and finite thickness corrections are included are considerably
 smaller than
what is found for two-dimensional layers and no mixing \cite{herb}.
 However, the resulting
energies are still  almost  a factor of two larger than what is found  in 
experiment \cite{eis}. The discrepancy is likely to be due
to disorder.

\begin{figure}[h]
\centerline{
  \epsfxsize=2.5in
  \epsffile{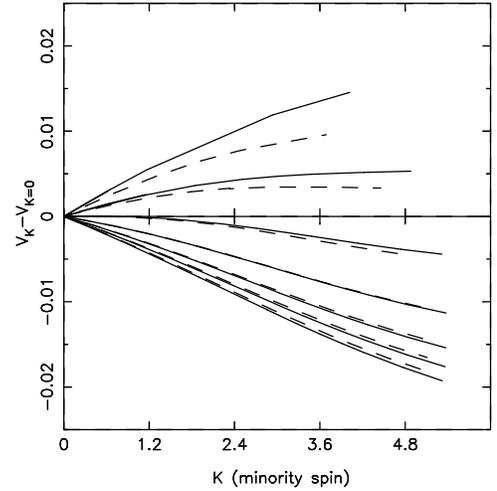}
}   
\vspace{0.2in}
\caption{ \label{first}
Energy difference between skyrmion and polarized quasihole at $\nu=3$;
the well width is  $1.73 \ell$ ; $\tilde g \rightarrow 0 $. 
The full lines correspond to the quasihole
and the dashed ones to the quasielectron case. The cyclotron energy range
 is between
  $100.0e^2/\kappa \ell$ for the lowest line and  
 $0.8 e^2/\kappa \ell$ for the uppermost,
each cyclotron energy being 50\% smaller than the previous higher one.   }
\end{figure}

\begin{figure}[h]
\centerline{
  \epsfxsize=2.5in
  \epsffile{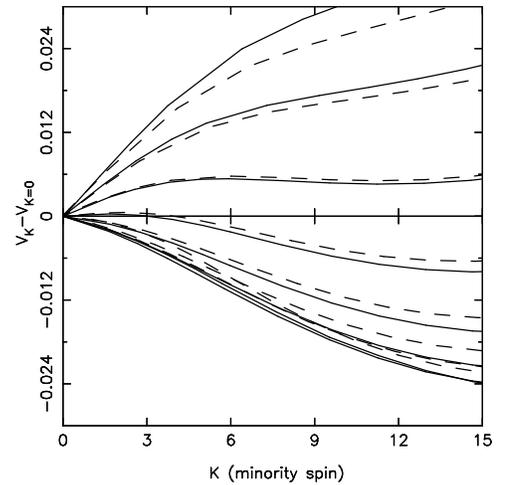}
}   
\vspace{0.2in}
\caption{ \label{second}
Energy difference between skyrmion and polarized quasihole at $\nu=5$;
the well width is  $1.73 \ell$  ; 
$\tilde g \rightarrow 0 $. The full lines correspond to the quasihole
and the dashed ones to the quasielectron case. 
The cyclotron  energies are the same as in
Fig. \ref{first}. } 
\end{figure}

 Figs. (\ref{first}) and (\ref{second}) present analogous results 
for $\nu=3$ and $\nu=5$.
Note the considerably smaller energy scales in these figures, indicating that
skyrmions can only be stable 
(if ever) for small values of $\tilde g$ \cite {cooper, herb2}.
It is apparent that  the introduction  of Landau level mixing 
 and screening destabilizes
 the skyrmion. Evidently, spin-polarized particles are better
 able to take advantage of the
admixture of higher Landau levels than skyrmions. 

\subsection {Effect of Finite Well Width}
\begin{figure}[h]
\centerline{
  \epsfxsize=2.5in
  \epsffile{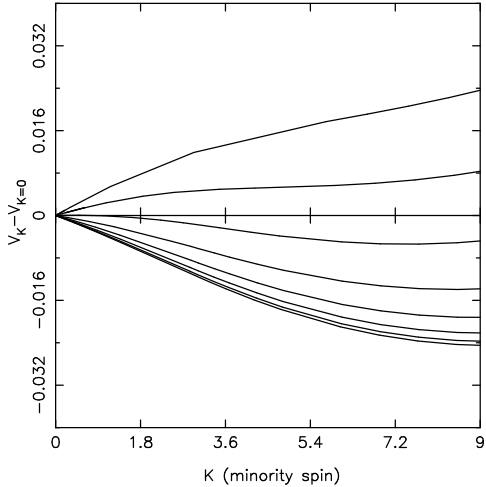}
}
\vspace{0.2in}
\caption{ \label{first173}
Energy difference between skyrmion and polarized quasihole at $\nu =3$  and
the well width of  $1.73 \ell$ ; $\tilde g \rightarrow 0 $.
The cyclotron energies are the same as in
Fig. \ref{first}. }
\end{figure}

\begin{figure}[h]
\centerline{
  \epsfxsize=2.5in
  \epsffile{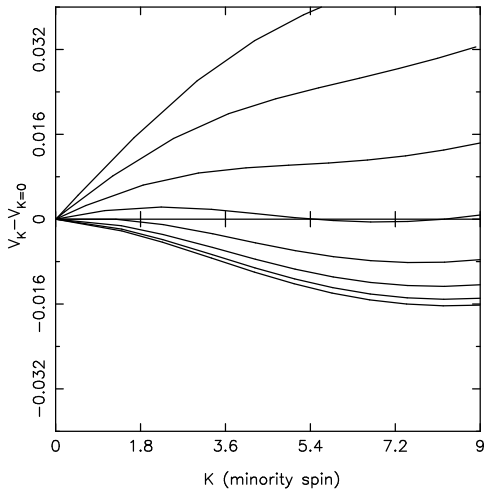}
}
\vspace{0.2in}
\caption{ \label{first100}
Energy difference between skyrmion and polarized quasihole at $\nu =3$  and
the well width of  $1.0 \ell$ ; $\tilde g \rightarrow 0 $.
The cyclotron energies are the same as in
Fig. \ref{first}. }
\end{figure}

\begin{figure}[h]
\centerline{
  \epsfxsize=2.5in
  \epsffile{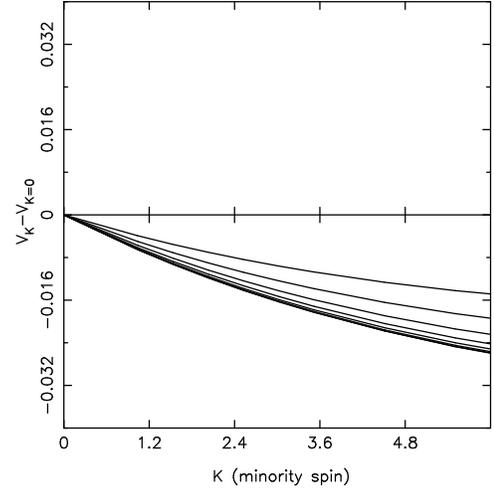}
}
 \vspace{0.2in}
\caption{   \label{sarma} Difference in energy between the skyrmionic and 
polarized quasihole solutions using Zhang-Das Sarma potential. The well width
is $1.73 \ell$ and $\nu = 3$. The cyclotron energies are the same as in
Fig. \ref{first}. }  
\end{figure}

Quasiparticle energies   depend on the well width, as illustrated in
 Figs. \ref{first173} and \ref{first100}.
As expected \cite{cooper, herb2}, we find that
for narrower wells the difference in energy is less favorable for the skyrmion
(Fig. \ref{first100}). Note  that the width used in Fig. \ref{first173}
is close to an experimentally reported value \cite{song}.
It is worth remarking at this juncture that a reasonably realistic model
of the electron-electron interaction with finite sample thickness corrections
is needed to obtain qualitatively correct results. Fig. \ref{sarma} 
shows that  the use of a simpler model
potential (Zhang and Das Sarma \cite{zhang})
\begin{eqnarray} \label{ZDS}
v_{ZdS}(\vec r - \vec r') =
\frac {1}{ \sqrt{|\vec r - \vec r'|^2 +w^2}}
\end{eqnarray}
which is 
commonly used in studying quantum Hall systems  
(see for example Refs. \citen{melik} and \citen{shankar})
gives substantially different results then those presented above 
(Fig. \ref{first}).
The principal difference between 
$v_C(r)$ and $v_{ZdS}(r)$ is the behavior at small r; 
the former diverges logarithmically, whereas the latter is regular.
Other divergent potentials give results consistent with Figs. \ref{first} and
  \ref{second}; it is likely that the  oversimplified behavior 
at short distances is responsible for the poor performance of  $v_{ZdS}(r)$
in this problem.
\subsection{Phase Diagram}
 Based on results in Figs. ~\ref{lowest}, ~\ref{first}, and ~\ref{second} 
we can construct the phase diagrams 
of skyrmion stability for  the filling factors of $\nu = 1,3,$ and $5$. 
Large $\omega _c$ and small 
$g_c$ is the region favoring the spin structured excitations. We see that   the 
region ``shrinks''
 as one moves to the higher filling  factors. Also, according to this 
calculation the breaking of symmetry between the quasihole and
 quasielectron excitations
upon lowering the cyclotron energy is quite spectacular 
 in the lowest Landau level,
whereas it plays no significant role  in the higher ones.
\begin{figure}[h]
\centerline{
  \epsfxsize=2.5in
  \epsffile{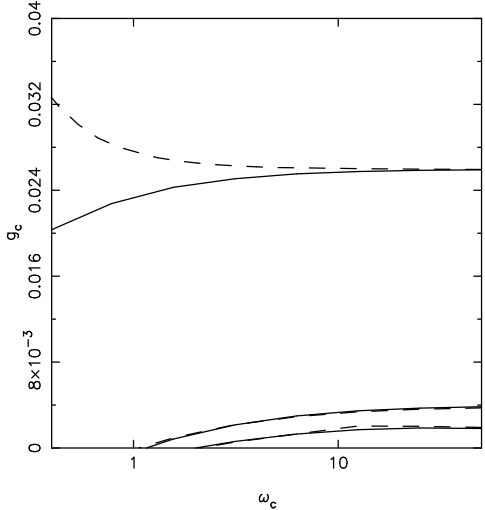}
}\vspace{0.2in}
\caption{\label{phased} 
Phase diagram of skyrmion stability. 
The dashed lines correspond to
the the spin  structured quasielectron case,
 whereas the full lines correspond the quasihole 
one. Top lines correspond to the case of $\nu =1$,  middle to $\nu =3$
and bottom to  $\nu =5$.
The scale on the abscissa is logarithmic. The width  is  $w = 1.73 \ell$.  }
\end{figure}
 In the paper by Song \emph{et al}.  ~\cite{song} 
 the reported excitation at the parameter
values of $w = 1.73 \ell,\omega _c = 0.56$,
 and $ \tilde g = 1.72 \times 10^{-2}  $  
falls well outside the boundary expected from the Hartree-Fock calculation.
\subsection {Effect of Impurities}
\begin{figure}[h]
\centerline{
  \epsfxsize=2.5in
  \epsffile{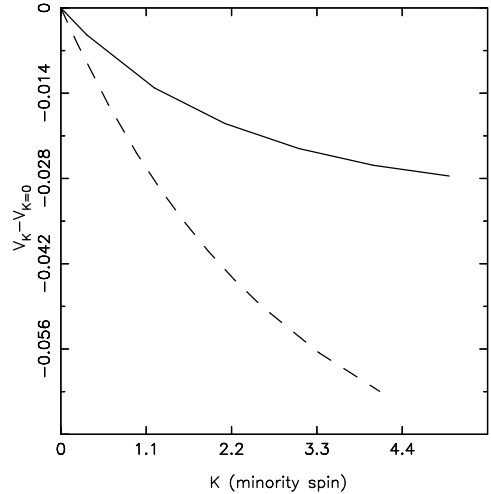}
}\vspace{0.2in}
\caption{\label{imp2nu1} Comparison of skyrmion stability in a system 
without an  impurity (dashed line)  and the one with impurity
at the distance of 2 magnetic lengths from the sample for  $\nu=1$ case.
Cyclotron energy is $100.0 e^2/\kappa \ell$ in both cases. }
\end{figure}

 It is tempting to speculate that inclusion of impurity
effects can stabilize the skyrmions at values of $g$ bigger than
allowed in a pure sample. 
  To test this idea we can include a simple model of an  impurity
in our calculation: a  point  charge (impurity) at a distance
$z_0$ above the central plane of the system. It  is replaced by an effective
non-uniform charge  density in the plane producing the same potential,
\begin{eqnarray}
\frac{1}{ |\vec r- (\vec r_0 + \vec z_0)| } = \int d^2r'
	 \rho _{eff}(\vec r')\frac{1}{ |\vec r- \vec r' | } .
\end{eqnarray}
 The effective density  can be found to be
\begin{eqnarray}
\rho _{eff}(\vec r)=\frac{1}{2\pi}\frac {z_0}{(z_0^2+R^2)^{3/2}}
\end{eqnarray}
for $z_0>0$.
Results with and without such an impurity are illustrated in Fig. \ref{imp2nu1}.
As may be seen, the impurity favors spin-polarized quasiparticle over skyrmion.
A similar result is expected for a short-range (e.g. delta-function) impurity
potential. Apparently  the simplest models of disorder are not likely to
explain the results of Ref. \citen{song}. It is probable  that more
complicated impurities (e.g. multiply charged or magnetic ones) could
stabilize  the small-spin  skyrmions at $\nu = 3$. However, in the
absence of data indicating such types of disorder in real samples,
 an investigation of this phenomenon  is left for future work.
\begin{figure}[h]
\centerline{
  \epsfxsize=2.5in
  \epsffile{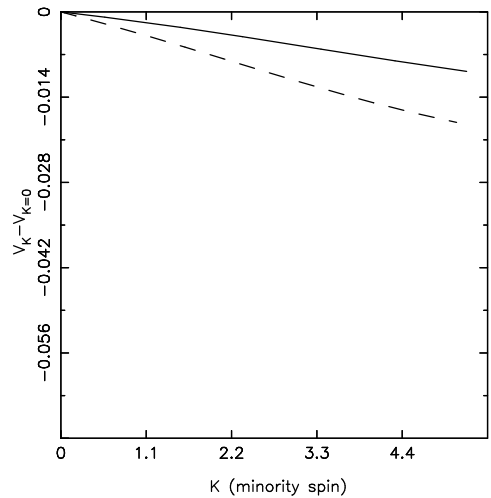}
}\vspace{0.2in}
\caption{\label{imp2nu3} The same as Fig. \ref{imp2nu1}
  for  $\nu=3$ case.  }
\end{figure}
\section{Conclusion}
In this paper we have presented a real-space method for computing 
Hartree-Fock states and energies of two-dimensional systems in
magnetic fields, appropriate for  systems with circular symmetry  in
which Landau level mixing may be important.  The method was  applied
to  compute the effects of  Landau level mixing and screening on skyrmion
states. It was found that   in most cases these tend to destabilize 
skyrmions, with a notable exception occurring for the case of the
quasielectron (antiskyrmion) around $\nu=1$. The calculations indicate
that  Hartree-Fock states  cannot account for the results of
 Ref. \citen{song} (reporting skyrmions at $\nu=3$). This is
in agreement with earlier studies where    Landau level mixing and screening
were not included.
\acknowledgements
This work was supported by NSF Grant Nos.DMR98-70681 and PHY94-07194,
the Research Corporation,  and the Center for Computational Sciences of
the University of Kentucky.


\begin{references}
\bibitem{lee} D.H. Lee and C.L. Kane, Phys. Rev. Lett. {\bf 64}, 1313
(1990).
\bibitem{sondhi} S.L.Sondhi, A.Karlhede, S.A.Kivelson, and E.H.Rezayi, 
    Phys. Rev. B {\bf 47}, 16419, (1993).
\bibitem{herb} H.A.Fertig, L.Brey, R.C\^{o}t\'{e},  and  A.H.MacDonald, 
 Phys. Rev. B {\bf 50},
  11018 (1994).
\bibitem{skyrme} T.H.R. Skyrme, Proc. Roy. Soc. {\bf A262}, 233 (1961).

\bibitem{rajaraman} R. Rajaraman, {\it Solitons and Instantons}
(North-Holland, Amsterdam, 1989).

\bibitem{polyakov} A.A. Belavin and A.M. Polyakov, JETP Lett. {\bf 22},
245 (1975).

\bibitem{barret} S.E.Barret, G.Dabbagh, L.N.Pfeiffer, and R.Tycko,
Phys. Rev. Lett. {\bf 74} 5112, 1995

\bibitem {brey} L.Brey,  H.A.Fertig, and R.C\^{o}t\'{e}, 
 and A.H.MacDonald Phys. Rev. Lett. {\bf 75}  2562 (1995).
 
\bibitem{eis}  A.Schmeller, J.P.Eisenstein, L.N.Pfeiffer, and K.W.West,
  Phys. Rev. Lett. {\bf 75}, 4290, 1995 	

\bibitem{aifer}  E.H. Aifer, B.B. Goldberg, and D.A. Broido, 
 Phys. Rev. Lett. {\bf 76}, 680 (1996)

\bibitem{cooper}    N.R.Cooper,  Phys. Rev. B {\bf 55}, 1934, (1997).


\bibitem{herb2} H.A. Fertig, L. Brey, R. C\^ot\'e, A.H. MacDonald, A. Karlhede,
and S.L. Sondhi, Phys. Rev. B {\bf 55}, 10671 (1997).


\bibitem {melik}   V.Melik-Alaverdian,  N.E.Bonesteel, and  G.Ortiz, 
  Phys. Rev. B {\bf 60}, R8501.

\bibitem{song}  Y.Song,   R.M.Goodson; K.Maranowski, A.C.Gossard,
       Phys. Rev. Lett. {\bf 82},2768, 1999



\bibitem{zhang} F.C. Zhang and S. Das Sarma, Phys. Rev. B {\bf 33}, 2903
(1986).

\bibitem{shankar} R.Shankar, cond-mat/9911288 
\end{references}
\end{document}